\documentclass[12pt,preprint]{revtex4}

\usepackage{amsmath}
\usepackage{graphicx}
\usepackage{amssymb}
\usepackage[none]{hyphenat}
\setcitestyle{numbers}

\begin{document}

\title{\bf{Quantum Optical Rotatory Dispersion}}

\author{Nora Tischler$^{1,2,3,\dagger,*}$, Mario Krenn$^{1,2}$, Robert Fickler$^{1,2,\ddagger }$, \\Xavier Vidal$^{3}$, Anton Zeilinger$^{1,2}$, and Gabriel Molina-Terriza$^{3}$}
\affiliation{{\footnotesize $^{1}$  Vienna Center for Quantum Science and Technology (VCQ), Faculty of Physics, University of Vienna, Boltzmanngasse 5, A-1090 Vienna,
Austria}}
\affiliation{{\footnotesize$^{2}$Institute for Quantum Optics and Quantum Information (IQOQI), Austrian Academy of Sciences, Boltzmanngasse 3, A-1090 Vienna,
Austria}}
\affiliation{{\footnotesize$^{3}$Department of Physics \& Astronomy, Centre for Engineered Quantum Systems, Macquarie University, NSW 2109, Sydney, Australia}}
\affiliation{{\footnotesize$^{\dagger}$present address: Centre for Quantum Dynamics, Griffith University, Brisbane 4111, Australia}}
\affiliation{{\footnotesize$^{\ddagger}$present address: Department of Physics and Max Planck Centre for
Extreme and Quantum Photonics, University of Ottawa, Ottawa, K1N 6N5, Canada}}
\affiliation{{\footnotesize$^{*}$Corresponding author. E-mail: nora.tischler@univie.ac.at}}


\begin{abstract}
{\vspace{0.9cm}\bf \noindent The phenomenon of molecular optical activity manifests itself as the rotation of the plane of linear polarization when light passes through chiral media. Measurements of optical activity and its wavelength dependence, optical rotatory dispersion, can reveal information about intricate properties of molecules, such as the 3D arrangement of atoms comprising a molecule. 
Given a limited probe power, quantum metrology offers the possibility to outperform classical measurements. This holds particular appeal when samples may be damaged by high powers, a potential concern for chiroptical studies.
Here we show the first experiment in which multi-wavelength polarization-entangled photon pairs are used to measure the optical activity and optical rotatory dispersion exhibited by a solution of chiral molecules. 
Our work paves the way for quantum-enhanced measurements of chirality, with potential applications in chemistry, biology, materials science, and the pharmaceutical industry. The scheme we employ for probing the wavelength dependence allows to surpass the information extracted per photon in a classical measurement, and can also be used for more general differential measurements.}
\end{abstract}

\maketitle       

\section*{INTRODUCTION}
\begin{figure}[ht]
	\centering
		\includegraphics[width=0.33\textwidth]{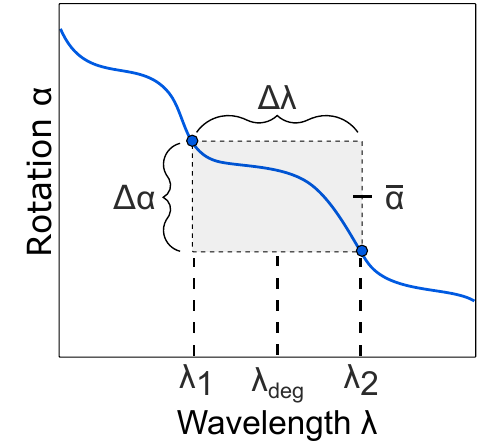}
	\caption{Illustration of the two quantities that were measured with two types of input states: The state $|\Phi_{\mathrm{in}}\rangle$ with correlated polarizations is sensitive to $\overline{\alpha}$, the mean of the optical rotations of the two wavelengths in question. In contrast, the state $|\Psi_{\mathrm{in}}\rangle$ with anti-correlated polarizations is sensitive to $\Delta \alpha$, the difference of the two optical rotations. The curve is a fictitious example for illustration purposes.}
	\label{fig:Quantities_illustration}
\end{figure}

The concept of chirality, which means that an object cannot be superposed onto its mirror image, pervades several fields of science, including chemistry, biology, physics, and materials science, and also plays a key role in the pharmaceutical industry \citep{Lough2002,Amabilino2009}.
One of the means by which we are able to probe the chirality of molecules is through its effect on light-matter interaction, manifesting as the phenomena of optical activity and circular dichroism \citep{Caldwell1971,Charney1979,Barron2004}.
Optical activity can be understood as circular birefringence, which results in a different phase for the two circular polarizations. 
Polarization-based phase estimation is amenable to techniques from quantum metrology that hold the possibility of beating the standard quantum limit, and has been demonstrated in a number of quantum metrology experiments in the past \citep{Kuzmich1999,Mitchell.2004,Sun2008,Higgins2009,Xiang2010,Afek2010,Wolfgramm2012,Xiang2013,Israel2014,Ono2013}. In one of these, Wolfgramm \emph{et al.} reported the measurement of Faraday rotation using a narrowband polarization NOON state  \citep{Wolfgramm2012}. This suggests that nonclassical resources could also be exploited in chiroptical techniques, given that Faraday rotation and molecular optical activity have a similar effect on light in spite of their different physical origins. However, a new approach is needed to accommodate the significance of the wavelength dependence in chiroptical techniques.
Addressing this challenge, we show how entangled photons can be used to study optical activity as a function of wavelength, with an enhancement in the information about multi-wavelength quantities that is extracted per photon. In particular, we will focus on the optical rotations for two wavelengths at a time, as illustrated in Fig.\ \ref{fig:Quantities_illustration}, and consider the mean and difference of the rotations, denoted by $\overline{\alpha}$  and $\Delta \alpha$. 

\section*{RESULTS}
\subsection*{Measurement scheme}
The straightforward classical approach to measuring optical rotations at two wavelengths would be to simply perform a measurement at each wavelength. Given the resource of two photons, this could be implemented with a separable photon pair where each photon has linear polarization, and measurements are performed in the horizontal-vertical (H-V) basis. We will use this as the classical benchmark, to which the quantum metrology scheme may be compared. 
For the quantum scheme of our experiment, two different types of biphoton input states are used. In both cases, the two photons are in separate paths, and may additionally have unequal wavelengths. The input states are the following polarization-entangled states:
\begin{eqnarray}
|\Phi_{\mathrm{in}}\rangle&\equiv&\frac{1}{\sqrt{2}}\left[|R,\lambda_{1}\rangle_{1}|R,\lambda_{2}\rangle_{2}+\mathrm{e}^{i\alpha_{0}}|L,\lambda_{1}\rangle_{1}|L,\lambda_{2}\rangle_{2}\right] \label{Eq:input1} \\
|\Psi_{\mathrm{in}}\rangle&\equiv&\frac{1}{\sqrt{2}}\left[|R,\lambda_{1}\rangle_{1}|L,\lambda_{2}\rangle_{2}+\mathrm{e}^{i\alpha_{0}}|L,\lambda_{1}\rangle_{1}|R,\lambda_{2}\rangle_{2}\right]. \label{Eq:input2}
\end{eqnarray}
Here, $|P,\lambda_1\rangle_1 |P',\lambda_{2}\rangle_{2}$ is shorthand for the two-photon composite system. $|P,\lambda_m\rangle_m$ symbolizes a photon with polarization $P$ and wavelength $\lambda_m$ in path $m$. The polarization is $L$ or $R$ for left or right circular polarization, which implies a helicity of 1 and -1, respectively. The phase $\alpha_0$ will be referred to as the bias phase, and the Bell states are recovered when $\alpha_0=n\pi$, $n\in \mathbb{Z}$.
Optical activity that is experienced by light passing through a solution of chiral molecules can be modeled as a unitary transformation of the form $U\left(\alpha\left(C,\lambda\right)\right)=\mathrm{exp}\left[-i\Lambda\alpha\left(C,\lambda\right)\right]$, where $\Lambda$ is the helicity of the light and $\alpha\left(C,\lambda\right)$ is the angle by which the linear polarization of single photons gets rotated, which is a function of the wavelength of the light $\lambda$ and of the concentration of the solution $C$. 
The measurement parameters are $\overline{\alpha}\equiv\frac{1}{2}\left(\alpha\left(C,\lambda_{1}\right)+\alpha\left(C,\lambda_{2}\right)\right)$  and $\Delta \alpha\equiv\left(\alpha\left(C,\lambda_{2}\right)-\alpha\left(C,\lambda_{1}\right)\right)$. Using the quantum input states and projective measurements in the H-V basis, the expectation values of the HH and VV coincidences are $\frac{1}{4}\left(1+\mathrm{cos}\left(\theta\right)\right)$, while the mixed coincidences HV and VH give $\frac{1}{4}\left(1-\mathrm{cos}\left(\theta\right)\right)$, where $\theta=\alpha_{0}-4\left(\overline{\alpha}\left(C,\lambda_{1},\lambda_{2}\right)\right)$ when $|\Phi_{\mathrm{in}}\rangle$ is used, and $\theta=\alpha_{0}+2\left(\Delta\alpha\left(C,\lambda_{1},\lambda_{2}\right)\right)$ for $|\Psi_{\mathrm{in}}\rangle$.

To assess the performance of the measurement schemes, we make use of the Fisher information (FI), which is a measure of the information about a parameter that can be extracted from the probe state with a given measurement procedure. It is
defined as $ I\left(\theta \right)\equiv \sum_i p(x_i|\theta) \left( \frac{\partial \mathrm{ln} p(x_i|\theta)}{\partial \theta} \right)^2$, where $p(x_i|\theta)$ is the probability of obtaining the measurement outcome $x_i$ given the parameter value of $\theta$, and the sum is taken over all measurement outcomes of the positive operator valued measure (POVM), for the case of our experiment the outcomes HH, HV, VH, and VV \citep{Kok2010}. As detailed in Section A of the Supplementary Materials, the two quantum input states distinguish themselves from the classical state through the interesting property that each is sensitive to one of the multi-wavelength quantities, $\overline{\alpha}$ or $\Delta \alpha$, while insensitive to the other. Using the state $|\Phi_{\mathrm{in}}\rangle$ provides a factor of two enhancement of the FI for the estimation of the mean rotation $\overline{\alpha}$ compared to the classical scheme, while $|\Psi_{\mathrm{in}}\rangle$ yields the factor two enhancement for the estimation of the rotation difference $\Delta \alpha$. Furthermore, the FI is equal to the quantum Fisher information (QFI), which is the FI optimized over all possible POVMs \citep{Giovannetti2011,Toth2014}. This shows that our projective measurement in the H-V basis is optimal. 

\subsection*{Experimental implementation}
\begin{figure}[htb]
	\centering
		\includegraphics[width=1\textwidth]{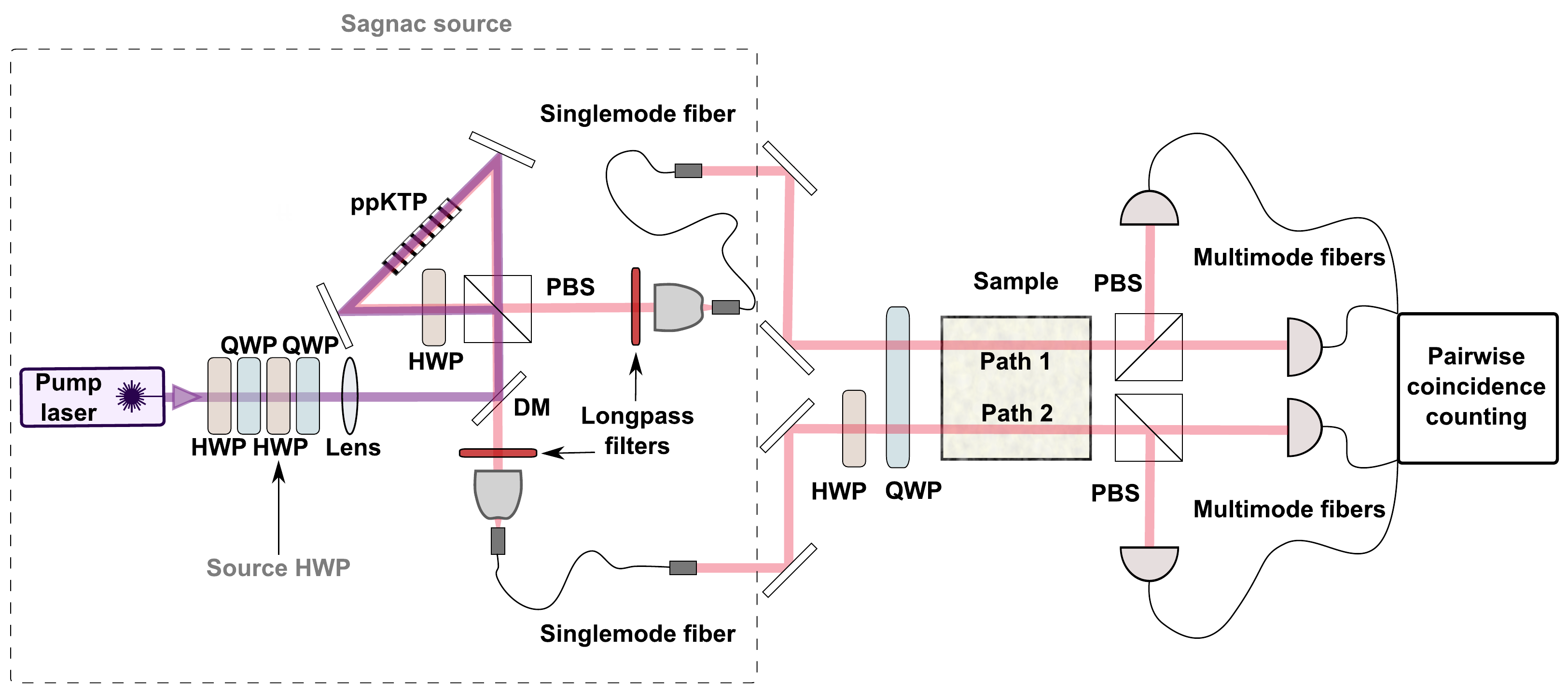}
	\caption{Schematic of the experimental set-up. Wavelength tunable polarization-entangled photon pairs in separate paths are created by type-II SPDC in a polarization-based Sagnac interferometer. The bias phase is controlled with the half waveplate (HWP) labeled ``Source HWP", which is in a set of four waveplates used to control the pump polarization (the first HWP controls the relative amplitude, and the ``Source HWP" the relative phase between H and V). The preparation of the desired polarization state of the photon pair is then completed by a HWP and quarter waveplate (QWP) before the sample (the HWP setting determines whether $|\Phi_{\mathrm{in}}\rangle$ of Eq.\ (\ref{Eq:input1}) or $|\Psi_{\mathrm{in}}\rangle$ of Eq.\ (\ref{Eq:input2}) is selected). Next, the photons propagate through the sample, which consists of a cuvette with a path length of 20 mm filled with either water or a sucrose solution. Afterwards, each photon is incident on a polarizing beam splitter (PBS), followed by multimode fiber coupling and detection with avalanche photodiodes (APDs). A coincidence logic enables the detection of the four types of coincidences in the H-V basis (HH, HV, VH, VV). DM: dichroic mirror, ppKTP: periodically poled potassium titanyl phosphate.}
	\label{fig:OA_project_setup}
\end{figure}

The experimental set-up with which we implemented the idea is sketched in Fig.\ \ref{fig:OA_project_setup}. A polarization-based Sagnac interferometer with type-II spontaneous parametric down-conversion (SPDC) is used to create a polarization-entangled biphoton state \citep{Kim2006,Fedrizzi2007}, from which $|\Phi_{\mathrm{in}}\rangle$ or $|\Psi_{\mathrm{in}}\rangle$ is prepared using fiber polarization controllers and waveplates. The bias phase $\alpha_0$ of Eqs.\ (\ref{Eq:input1}) and (\ref{Eq:input2}) is directly controlled with one of the half waveplates (HWP) that are used to prepare the polarization of the pump beam, while the wavelengths of the photons are tuned with the temperature of the nonlinear down-conversion crystal to yield a range between 800 and 819 nm. More details on the spectral properties are provided in Section B of the Supplementary Materials. 
In order to demonstrate the technique we used sucrose solutions. While sucrose solutions are not delicate, the choice of wavelength would be important for potential applications in biology since the damage of biological samples, as well as losses, are wavelength-dependent. 
In these matters, near infrared light tends to strike a favorable balance between competing unwanted effects \citep{Taylor2014}. 
The light-matter interaction takes place as both photons travel through the same sample, which provides a path length of 20 mm. The content of the cuvette is either water or a sucrose solution with a concentration $0.200\pm 0.002$ g/ml or $0.400\pm0.008$ g/ml. 
After the cuvette, a projective measurement of the photon pair in the H-V basis is performed, using two polarizing beam splitters and four APDs. 
Further technical details about the experimental set-up can be found in the Methods section.

\subsection*{Calibration measurements}

In preparation for the measurements of optical activity and optical rotatory dispersion of sucrose, a calibration was performed to determine the mapping from photon coincidence counts to phases, and to assess the experimental sensitivity. Instead of using an optically active sample, the calibration was based on polarization control through HWPs in the set-up, because this enabled a straightforward and wide tunability of the phases. The mapping was obtained through recording the coincidence counts as a function of the bias phase $\alpha_0$. The experimental visibilities for the different wavelength settings of the experiment range from 91.4\% to 93.6\%. 

\begin{figure}[htb!]
	\centering
		\includegraphics[width=0.9\textwidth]{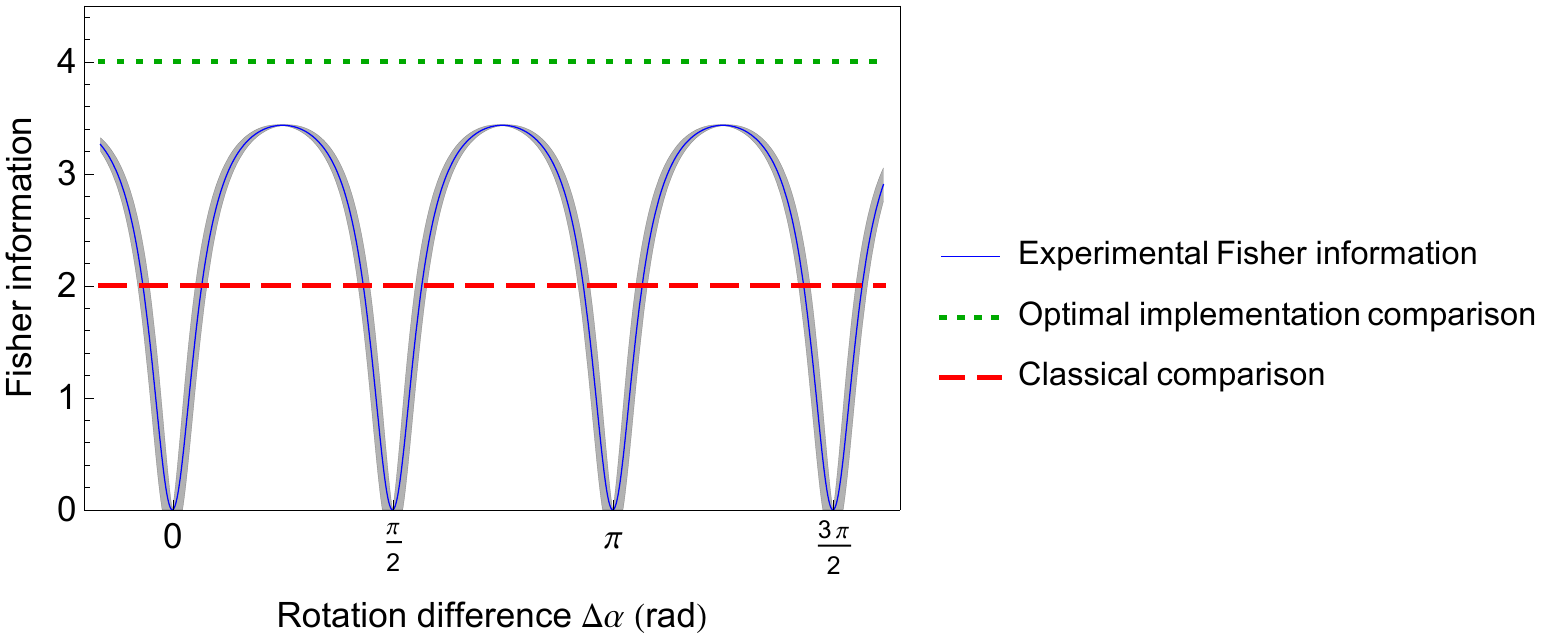}
	\caption{Comparison of the Fisher information as a function of the rotation difference for different lossless cases. The experimental FI inferred from the fitted sinusoidal curves of the calibration measurements for $|\Psi_{\mathrm{in}}\rangle$, obtained with a HWP controlling the bias phase and converted to the equivalent values of $\Delta \alpha$, is shown as the solid blue line (Section C of the Supplementary Materials). The narrow lighter blue shaded region represents the uncertainty ($\pm$ one standard deviation), as obtained from the uncertainty of the fit parameters with standard error propagation. The inferred experimental FI is presented alongside the FI from an ideal implementation of the quantum measurement scheme (green dotted  line), and the FI corresponding to an ideal classical measurement with linearly polarized photons (red dashed line). }
	\label{fig:FI}
\end{figure}

The fitted sinusoidal calibration curves of the four types of coincidences from a measurement using $|\Psi_{\mathrm{in}}\rangle$ at the degenerate wavelength (see Section C of the Supplementary Materials) were used to infer the experimental FI. Fig.\ \ref{fig:FI} shows a comparison of the FI as a function of the rotation difference for three cases: The case of an ideal implementation of the proposed scheme with the entangled state $|\Psi_{\mathrm{in}}\rangle$, the case of an ideal implementation of the classical scheme, and the inferred FI for the experimental implementation. The inferred experimental FI accounts for all the experimental details that lead to a reduced visibility, but not for the losses, because it is based on data post-selected for coincidences. Due to the visibilities being below 100\%, the experimental FI  possesses a dependence on the value of the rotation difference and does not reach the ideal value for entangled photons. Nonetheless, for any given rotation difference the bias phase can be used to shift the curve horizontally such that the experimental FI surpasses the classical FI. As the subsequent optical activity measurements involved small polarization rotations, the bias phase was adjusted to maximize the sensitivity for these main measurements. 

\subsection*{Optical rotatory dispersion measurements}
\begin{figure}[ht]
	\centering
		\includegraphics[width=0.95\textwidth]{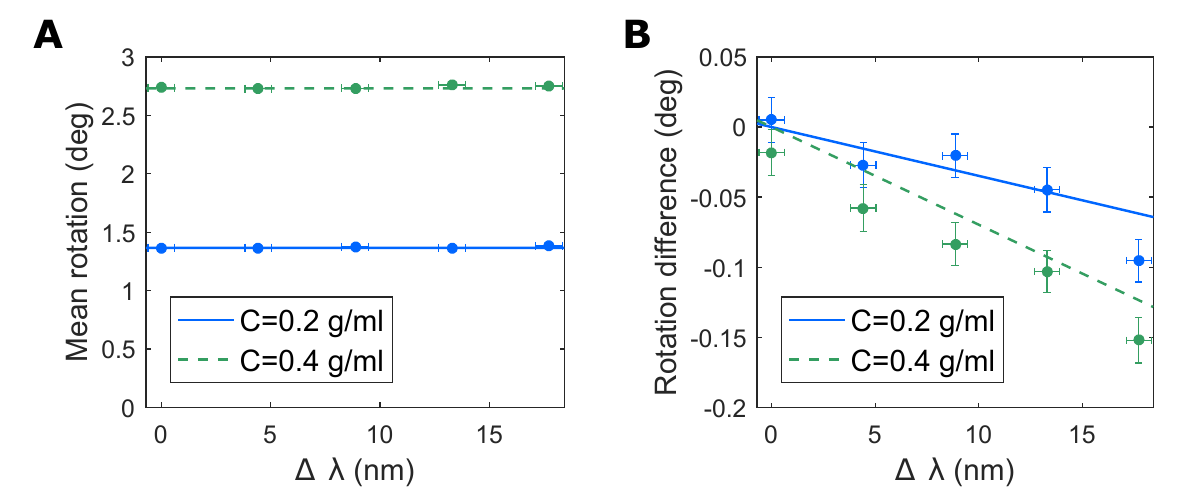}
	\caption{Comparison of experimental optical activity measurement results (data points) for sucrose with predictions (lines). ({\bf A}) depicts the mean rotations obtained from using $|\Phi_{\mathrm{in}}\rangle$, and ({\bf B}) the rotation differences obtained from using $|\Psi_{\mathrm{in}}\rangle$, for wavelengths of 809.7 $\pm \frac{\Delta \lambda}{2}$ nm. The results for a concentration of 0.4 g/ml are given by the dashed green line, and for 0.2 g/ml by the solid blue line. The horizontal error bars indicate $\pm$ one standard deviation of the wavelength difference between the photons in a pair, estimated based on the spectrum shown in Fig.\ S1B of the Supplementary Materials. The vertical error bars show $\pm$ one standard error of the mean, estimated from the standard deviation of the experimentally obtained angles. For the cases where the error bars are not clearly visible, the intervals are smaller than the markers. The measurements were taken at a temperature of $19^{\circ}$C and with a setting of the bias phase to maximize the FI.}
	\label{fig:OptActResults3}
\end{figure}

For the optical activity study, we measured the mean and difference of the optical rotations at five pairs of wavelengths, which are at varying distances from the degenerate wavelength 809.7 nm, and conducted these measurements using two different sucrose concentrations. A sequence of seven-minute sets was taken, which consist of one-second acquisitions recording the four types of coincidences. For each of the two input states, each of the two sucrose concentrations, and each of the five crystal temperatures used, one data set was taken with water, and one with a sucrose solution. The measurement results are shown in Fig.\ \ref{fig:OptActResults3}. Part A contains the results for mean rotations and B for the rotation differences. Overlaid are the curves predicted by an empirical model (Section D of the Supplementary Materials). Due to the optical activity being proportional to the concentration, for both means and differences the values for the higher concentration $0.4 $ g/ml are expected to be a factor of two larger than for the lower concentration of $0.2$ g/ml. The reason behind the mean rotations being approximately constant and the rotation differences being linear functions is that the optical rotatory dispersion curve is nearly linear within the spectral region that is probed (Section D of the Supplementary Materials). 

The graphs show the expected overall behavior, and the model notably has no free fitting parameters. The systematic deviation in Fig.\ \ref{fig:OptActResults3}B for the concentration of 0.4 g/ml might be due to an unintended difference between conditions, e.g.\ concentrations or other parameters that affect the rotations, for the two paths in the cuvette.

Apart from demonstrating the possibility of directly accessing the mean and difference rotation, it is also interesting to compare the experimental uncertainties with what could have ideally been achieved in the classical measurement. We calculated the classically achievable uncertainties using the Cram\'{e}r-Rao bound (Section A of the Supplementary Materials) for the given number of coincidences that were detected in the measurement sets belonging to the data points, and compared them to the corresponding experimental values of the standard error of the mean. For the 20 data points of Fig.\ \ref{fig:OptActResults3}, the experimental standard error of the mean divided by the optimal uncertainty for the classical measurement is $0.77\pm 0.02$. 

\section*{DISCUSSION}
This experiment is the first to use entangled photon states for quantum-enhanced  differential measurements between different wavelengths. Our application of the method to the phenomenon of molecular optical activity, where the wavelength dependence is important, enabled us to measure the key quantity of optical rotatory dispersion.
Another interesting feature of the experiment is the use of distinguishable photons, as it is in contrast to the majority of optical quantum metrology experiments. One of the exceptions was the 2013 experiment by Bell \emph{et al.}, which also employed multi-wavelength entangled photons \citep{Bell2013}. There, the probe state was produced by four-wave mixing, and it was used to measure an optical path length in an interferometer, which is comparable to the part of our optical activity experiment where $|\Phi_{\mathrm{in}}\rangle$ is used to measure the mean rotation. By exploiting the distinguishability of the photons, both experiments avoid the limitation in recording some of the measurement outcomes that occurs in schemes where indistinguishable photon pairs are split only probabilistically before measurement by non-photon number resolving detectors. 
 
In our study of optical activity, measuring the difference between two wavelengths was of particular interest. However, the same method can also be used for more general differential measurements. As further examples involving optical activity, the current set-up can be adapted to compare the concentrations or enantiomeric purities of two solutions by choosing the photons to be at the same wavelength and simply performing the differential measurement across two different solutions.
 In magnetometry, the possibility of probing spatial field gradients with macroscopic singlet states has been established \citep{Toth2014,Urizar-Lanz2013}. 
An extension to states involving more photons is also possible for optical measurements. Photonic singlet states of higher photon numbers produced in parametric down-conversion have been proposed for the measurement of symmetry-breaking effects \citep{Cable2010}. While singlet states are invariant to rotations about any axis, another option for an extension to higher photon numbers is the family of states constructed as a superposition of the eigenstates corresponding to the minimum and maximum eigenvalues of the generator of the unitary transformation \citep{Giovannetti2011,Giovannetti2006,Zwierz2012}. Similar states have also been considered in the context of quantum clock synchronization \citep{Jozsa2000,Zhang2013}.

\section*{MATERIALS AND METHODS} \nonumber \label{Methods}
For the SPDC process a narrowband continuous wave laser (Ondax SureLock diode laser) with a wavelength of 404.85 nm and a power of 2.5 mW was used as the pump. The nonlinear crystal was a 15 mm long periodically poled potassium titanyl phosphate (ppKTP) crystal, phase matched for near degenerate down-conversion at the pump wavelength. 

When transitioning between different wavelength settings, the set-up was left unchanged, except for the required shift in the crystal temperature, which was controlled with an oven, and an adjustment of the HWP that controls the bias phase. This adjustment was found to be necessary in order to maintain a constant bias phase, most likely due to dispersive birefringent elements, such as waveplates, in the set-up. Typical background-corrected visibilities of the biphoton state at the output from the Sagnac interferometer in the diagonal-antidiagonal basis were $97.99\pm 0.08\%$. 

After coupling into singlemode fibers, unwanted polarization transformations in the fibers and subsequent mirrors were compensated with fiber polarization controllers. As a last step before incidence on the sample, the state $\frac{1}{\sqrt{2}}\left[|H,\lambda_{1}\rangle_{1}|V,\lambda_{2}\rangle_{2}-\mathrm{e}^{i\alpha_{0}}|V,\lambda_{1}\rangle_{1}|H,\lambda_{2}\rangle_{2}\right]$ from the quantum source was transformed to the desired input states $|\Phi_{\mathrm{in}}\rangle$ and $|\Psi_{\mathrm{in}}\rangle$ using one of two settings of a HWP for one of the paths, and a QWP for both paths.

For the sample, we used water HPLC grade from Hartenstein GmbH, and for the sucrose solutions D(+)-Sucrose pure Ph. Eur., NF from Hartenstein GmbH was dissolved in this water.

Although for clarity the two photonic paths are drawn horizontally beside each other in Fig.\ \ref{fig:OA_project_setup}, the approximately 1 cm separation in our implementation was  in fact vertical due to the dimensions of the cuvette. 

A number of measures were taken to stabilize the photon counts and polarizations within the experiment. In particular, the section of the set-up containing the biphoton source was enclosed and insulated to minimize vibrations and temperature fluctuations. The temperature of the singlemode fibers was also stabilized with insulating tape. 

Since standard equipment was used, there were considerable losses in the experiment: Without accounting for the singlemode fiber coupling of the photons in the Sagnac source and for the detection efficiency of the APDs, the single photon transmission through the set-up was typically 46-56\%. However, overall losses in the experiment could be significantly reduced by using high efficiency detectors and anti-reflection coated elements throughout. An improved efficiency would be required to enable a quantum enhanced precision when accounting for all of the photons used and not just the detected photons.

For the optical activity measurements that make up the results of Fig.\ \ref{fig:OptActResults3}, the average number of coincidences recorded per second is $1.837\times 10^4$ and the total number of coincidences that contribute to each of the 40 seven-minute data sets ranges between $7.287\times 10^6$  and $8.203\times 10^6$. An analysis of the noise, which shows that the coincidence counts are close to Poisson distributed, is provided in Section E of the Supplementary Materials. 

As accidental coincidences do not contribute to the sinusoidal signal with respect to the measured phase, they are one of the reasons for the non-unity fringe visibility, although the primary reason are imperfections in the state preparation. Accidentals were not subtracted as part of the analysis of the optical activity measurement data, because their effect is taken into account by the calibration curves (Section C of the Supplementary Materials). 

\nocite{Crespi2012}
\nocite{Lowry1924}

\section*{SUPPLEMENTARY MATERIALS} \nonumber \label{SIlist}
\noindent Section A: Fisher information calculations.\\
Section B: Spectral characterization of the biphoton state.\\
Section C: Calibration curves.\\
Section D: Optical activity predictions.\\
Section E: Measurement noise.

\bibliography{OvercompleteReferences}

\begin{thebibliography}{10}

\bibitem{Lough2002}
W.~J. Lough, I.~W. Wainer, eds., {\it {Chirality in natural and applied
  science}\/} (Blackwell Science Ltd, 2002).

\bibitem{Amabilino2009}
D.~B. Amabilino, ed., {\it {Chirality at the nanoscale: Nanoparticles,
  surfaces, materials and more}\/} (Wiley, 2009).

\bibitem{Caldwell1971}
D.~J. Caldwell, H.~Eyring, {\it {The theory of optical activity}\/}
  (Wiley-Interscience, 1971).

\bibitem{Charney1979}
E.~Charney, {\it {The molecular basis of optical activity: Optical rotatory
  dispersion and circular dichroism}\/} (Wiley, 1979).

\bibitem{Barron2004}
L.~D. Barron, {\it {Molecular light scattering and optical activity}\/}
  (Cambridge University Press, 2004).

\bibitem{Kuzmich1999}
A.~Kuzmich, L.~Mandel, {\it Quantum Semiclass.\ Opt.\/} {\bf 10}, 493 (1998).

\bibitem{Mitchell.2004}
M.~W. Mitchell, J.~S. Lundeen, A.~M. Steinberg, {\it Nature\/} {\bf 429}, 161
  (2004).

\bibitem{Sun2008}
F.~W. Sun, {\it et~al.\/}, {\it EPL\/} {\bf 82}, 24001 (2008).

\bibitem{Higgins2009}
B.~L. Higgins, {\it et~al.\/}, {\it New J.\ Phys.\/} {\bf 11}, 073023 (2009).

\bibitem{Xiang2010}
G.~Y. Xiang, B.~L. Higgins, D.~W. Berry, H.~M. Wiseman, G.~J. Pryde, {\it
  Nature Photon.\/} {\bf 5}, 43 (2011).

\bibitem{Afek2010}
I.~Afek, O.~Ambar, Y.~Silberberg, {\it Science\/} {\bf 328}, 879 (2010).

\bibitem{Wolfgramm2012}
F.~Wolfgramm, C.~Vitelli, F.~A. Beduini, N.~Godbout, M.~W. Mitchell, {\it
  Nature Photon.\/} {\bf 7}, 28 (2013).

\bibitem{Xiang2013}
G.~Y. Xiang, H.~F. Hofmann, G.~J. Pryde, {\it Sci.\ Rep.\/} {\bf 3}, 2684
  (2013).

\bibitem{Israel2014}
Y.~Israel, S.~Rosen, Y.~Silberberg, {\it Phys.\ Rev.\ Lett.\/} {\bf 112},
  103604 (2014).

\bibitem{Ono2013}
T.~Ono, R.~Okamoto, S.~Takeuchi, {\it Nat.\ Commun.\/} {\bf 4}, 2426 (2013).

\bibitem{Kok2010}
P.~Kok, B.~W. Lovett, {\it {Optical quantum information processing}\/}
  (Cambridge University Press, 2010).

\bibitem{Giovannetti2011}
V.~Giovannetti, S.~Lloyd, L.~Maccone, {\it Nature Photon.\/} {\bf 5}, 222
  (2011).

\bibitem{Toth2014}
G.~T\'{o}th, I.~Apellaniz, {\it J.\ Phys.\ A\/} {\bf 47}, 424006 (2014).

\bibitem{Kim2006}
T.~Kim, M.~Fiorentino, F.~N.~C. Wong, {\it Phys.\ Rev.\ A\/} {\bf 73}, 012316
  (2006).

\bibitem{Fedrizzi2007}
A.~Fedrizzi, T.~Herbst, A.~Poppe, T.~Jennewein, A.~Zeilinger, {\it Opt.\
  Express\/} {\bf 15}, 15377 (2007).

\bibitem{Taylor2014}
M.~A. Taylor, W.~P. Bowen, {\it Phys.\ Rep.\/} {\bf 615}, 1 (2016).

\bibitem{Bell2013}
B.~Bell, {\it et~al.\/}, {\it Phys.\ Rev.\ Lett.\/} {\bf 111}, 093603 (2013).

\bibitem{Urizar-Lanz2013}
I.~Urizar-Lanz, P.~Hyllus, I.~L. Egusquiza, M.~W. Mitchell, G.~T\'{o}th, {\it
  Phys.\ Rev.\ A\/} {\bf 88}, 013626 (2013).

\bibitem{Cable2010}
H.~Cable, G.~A. Durkin, {\it Phys.\ Rev.\ Lett.\/} {\bf 105}, 013603 (2010).

\bibitem{Giovannetti2006}
V.~Giovannetti, S.~Lloyd, L.~Maccone, {\it Phys.\ Rev.\ Lett.\/} {\bf 96},
  010401 (2006).

\bibitem{Zwierz2012}
M.~Zwierz, C.~A. P\'{e}rez-Delgado, P.~Kok, {\it Phys.\ Rev.\ A\/} {\bf 85},
  042112 (2012).

\bibitem{Jozsa2000}
R.~Jozsa, D.~S. Abrams, J.~P. Dowling, C.~P. Williams, {\it Phys.\ Rev.\
  Lett.\/} {\bf 85}, 2010 (2000).

\bibitem{Zhang2013}
Y.-L. Zhang, Y.-R. Zhang, L.-Z. Mu, H.~Fan, {\it Phys.\ Rev.\ A\/} {\bf 88},
  052314 (2013).

\bibitem{Crespi2012}
A.~Crespi, {\it et~al.\/}, {\it Appl.\ Phys.\ Lett.\/} {\bf 100}, 233704
  (2012).

\bibitem{Lowry1924}
T.~M. Lowry, E.~M. Richards, {\it J.\ Chem.\ Soc., Trans.\/} {\bf 125}, 2511
  (1924).

\end{thebibliography}

\bibliographystyle{Science}


\vspace{1cm}

\noindent \textbf{Acknowledgments}: The authors thank Alexander Bismarck and the Polymer and Composite Engineering (PaCE) group of the University of Vienna for access to laboratory equipment. N.T. would further like to thank William Plick, Dominic Berry, and Alexei Gilchrist for helpful discussions.  

\noindent The project was supported by the Austrian Academy of Sciences (\"{O}AW), the Austrian Science Fund (FWF) with SFB F40 (FOQUS), the Australian Research Council's Centre of Excellence for Engineered Quantum
Systems (EQuS) grant no.\ CE110001013, and the Australian Research Council's Discovery
Project (DP160103332). G.\ M.-T.\ was also supported through an Australian Research Council Future Fellowship.

\end{document}